\documentclass[final,3p,11pt]{elsarticle}

\usepackage{latexsym}
\usepackage{graphicx}
\usepackage{bbm}
\usepackage{amsmath,amssymb, mathrsfs}
\usepackage[latin1]{inputenc}
\usepackage{mciteplus}

\newcommand{\ktt}{k_\perp}
\newcommand{\ptt}{p_\perp}
\newcommand{\qtt}{q_\perp}

\def\p{{\boldsymbol p}}
\def\q{{\boldsymbol q}}

\newcommand{\qs}{Q_\mathrm{s}}
\newcommand{\lqcd}{\Lambda_{_{\rm QCD}}}
\newcommand{\as}{{\alpha_\mathrm{s}}}

\newcommand{\xt}{\boldsymbol{x}_\perp}
\newcommand{\yt}{\boldsymbol{y}_\perp}
\newcommand{\pt}{\boldsymbol{p}_\perp}
\newcommand{\qt}{\boldsymbol{q}_\perp}
\newcommand{\kt}{\boldsymbol{k}_\perp}
\newcommand{\nc}{{N_\mathrm{c}}}

\newcommand{\da}{{d_\mathrm{A}}}

\newcommand{\nr}[1]{(\ref{#1})} 
\newcommand{\ud}{\mathrm{d}}
\newcommand{\fig}{fig.~}

\newcommand{\eq}{eq.~}

\newcommand{\nf}{{N_\mathrm{F}}}
\newcommand{\nb}{{N_\mathrm{B}}}

\newcommand{\rmF}{{\mathrm{F}}}
\newcommand{\rmB}{{\mathrm{B}}}

\newcommand{\nref}{{N_\mathrm{R}}}

\begin{document}

\begin{frontmatter}

\title{Multiplicity distributions and long range
rapidity correlations}
\author[jyu,hip]{T. Lappi}
\address[jyu]{Department of Physics, %
 P.O. Box 35, 40014 University of Jyv\"askyl\"a, Finland}
\address[hip]{
Helsinki Institute of Physics, P.O. Box 64, 00014 University of Helsinki, Finland
}

\begin{abstract}
The physics of the initial conditions of heavy ion collisions is
dominated by the nonlinear gluonic interactions of QCD. These lead to
the concepts of parton saturation and the
Color Glass Condensate (CGC). We discuss recent progress
in calculating multi-gluon correlations in this framework, prompted 
by the observation that these correlations are in fact easier to compute in a 
dense system (nucleus-nucleus) than a dilute one (proton-proton).
\end{abstract}


\end{frontmatter}

\section{Introduction}
Bulk particle production in relativistic collisions around midrapidity 
originates from
small $x$ degrees of freedom, predominantly gluons,
 in the wavefunctions of the colliding hadrons or nuclei.
At large energies these gluons form a dense system 
characterized by a \emph{saturation scale} $\qs$.
The degrees of freedom with $p_T \lesssim \qs$ are fully nonlinear Yang-Mills 
fields with large field
strength $A_\mu \sim 1/g$ and occupation numbers $\sim 1/\as$; they can 
therefore be understood
as classical fields radiated from the large $x$ partons. 
Because of their large longitudinal momentum, the large $x$ degrees 
of freedom are effectively ``frozen'' during the interaction. They 
can be described as random color charges drawn from
a classical probability distribution $W_y[\rho]$ that 
depends on the rapidity cutoff $y=\ln 1/x$ separating the
large and small $x$ degrees of freedom. The dependence of
$W_y[\rho]$ on  $y$ is described by a Wilsonian 
renormalization group equation known by the acronym JIMWLK.
Note that while  this description is inherently 
nonperturbative, it is still based on a weak coupling argument, because the 
classical approximation requires 
$\as(\qs)$ to be small and therefore $\qs \gg \lqcd$.
The Color Glass Condensate (CGC, for reviews 
see~\cite{Iancu:2003xm,*Weigert:2005us,*Gelis:2010nm,*Lappi:2010ek}) 
is a systematic effective theory (effective because the large $x$ part
 of the wavefunction is  integrated out) description of the classical 
small $x$ degrees of freedom.

The term glasma~\cite{Lappi:2006fp}  refers to 
the coherent, classical field configuration resulting from the collision of 
two such objects CGC. The glasma fields are initially longitudinal, 
whence the ``glasma flux tube''~\cite{Dumitru:2008wn,Gavin:2008ev} picture. More importantly
for computing multigluon correlations, they are boost invariant
(to leading order in the QCD coupling) and depend on the 
transverse coordinate with a characteristic correlation 
length $1/\qs$. There are several signals in the RHIC 
data~\cite{Putschke:2007mi,*Daugherity:2008su,*Wenger:2008ts,Abelev:2009dq} that point to strong 
correlations originating from the initial stage of the collision. 
The glasma fields provide a natural framework for understanding
these effects, although much work is still left to do in understanding
the interplay with purely geometrical effects from the fluctuating
positions of the nucleons in the colliding 
nuclei~\cite{Konchakovski:2008cf,*Bzdak:2009xq,*Alver:2010gr,*Alver:2010dn,Lappi:2009vb}.

We shall first describe some general observations on computing
multigluon correlations in the glasma, arguing in 
Sec.~\ref{sec:multig} that they are in some sense simpler
to compute in a collision of two dense, saturated nuclear wavefunctions
than in the dilute limit (see  
Ref.~\cite{Gelis:2008rw,*Gelis:2008ad} for a more formal discussion).
We shall then, in Sec.~\ref{sec:negbin}
discuss one application of these ideas to computing
the multiplicity distribution of gluons in the collision before
moving to the leading $\ln 1/x$ rapidity dependence of the correlation
in Sec.~\ref{sec:rapdep}.

\section{Multigluon correlations in the glasma}
\label{sec:multig}

The gluon fields in the glasma are nonperturbatively strong, $A_\mu \sim 1/g$. 
This means that the gluon multiplicity  is $N \sim 1/\as$. For
a fixed configuration of the classical color sources it is well known that
the multiplicity distribution of produced gluons is Poissonian, 
i.e. $\langle N^2 \rangle - \langle N \rangle^2 = \langle N \rangle$. 
In this case the correlations and fluctuations in the gluon multiplicity
are all quantum effects that appear only starting from the one-loop level, 
i.e. suppressed by a power of the coupling constant $\as$.
The computation in the CGC framework does not end here, however. To calculate
the moments of the gluon multiplicity distribution one must first 
calculate the gluon spectra for fixed configuration of the color charges 
$\rho$ and then average over the probability distribution
$W_y[\rho(\xt)]$.
For the $n$th moment of the multiplicity distribution, i.e. an $n$-gluon correlation,
the leading order result is
\begin{equation}
\left\langle
\frac{\ud N}{\ud^3\p_1}
\cdots 
\frac{\ud N}{\ud^3\p_n}
\right\rangle 
= \left[ \,
\int\limits_{[\rho]} 
W\big[\rho_1(y)\big]  W\big[\rho_2(y)\big]
\left.  \frac{\ud N}{\ud^3\p_1}\right|_{_{\rm LO}}
\cdots
\left.  \frac{\ud N}{\ud^3\p_n}\right|_{_{\rm LO}}
\right],
\end{equation}
where the subsrcipt ``LO'' refers to the single gluon spectrum evaluated 
from the classical field configuration corresponding to a fixed
configuration of color charges.
This averaging, even after the subsequent subtraction of the
appropriate disconnected contributions, introduces a correlation 
already at the leading order in $\as$, i.e. enhanced by 
an additional $1/\as$ compared to the quantum correlations.
A natural example is the negative binomial distribution that 
we shall discuss below, whose variance is 
$\langle N^2 \rangle - \langle N \rangle^2 = 
\langle N \rangle^2/k + \langle N \rangle$. 
One must emphasize here that although these contributions arise
as formally classical correlations in the effective theory that is the
CGC, they are physically also quantum effects, where the weak coupling is 
compensated by a large logarithm of the energy that has been
resummed into the probability distribution $W_y[\rho(\xt)]$. In this
sense the leading correlations are present already in the wavefunctions
of the colliding objects.

\begin{figure}
\centerline{\resizebox{0.7\textwidth}{!}{
\includegraphics[height=5cm]{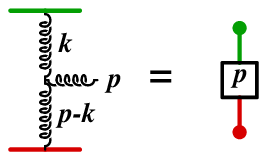}
\rule{3cm}{0pt}
\includegraphics[height=5cm]{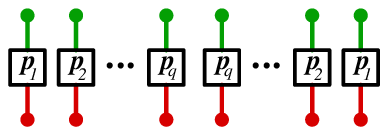}
}}
\caption{Left: Building block, Lipatov vertex coupled to two sources.
Right: combinatorics of the sources. The combinatorial problem is
to connect the dots on the upper and lower side (left- and right moving 
sources) pairwise.
}\label{fig:lipatov} \label{fig:combinat}
\end{figure}

\section{Multiplicity distribution}
\label{sec:negbin}

We can then apply this formalism to the calculation of the probability 
distribution of the number of gluons in the glasma~\cite{Gelis:2009wh}.
We shall assume the ``AA'' power counting of sources that
are parametrically strong in $g$, but nevertheless work to the lowest 
nontrivial order in the color sources. Formally this would correspond
to a power counting $\rho \sim g^{\varepsilon -1}$ with a small 
$\epsilon > 0$. In this limit, as we have discussed, the dominant
contributions to multiparticle correlations come from diagrams that
are disconnected for fixed sources and become connected only
after averaging over the color charge configurations. 
The corresponding two gluon correlation
function was computed in Ref.~\cite{Dumitru:2008wn} and
 generalized to a three gluons in Ref.~\cite{Dusling:2009ar}.
We shall here sketch the derivation ~\cite{Gelis:2009wh} 
of the general $n$-gluon  correlation in this simplified limit.

\begin{figure}
\begin{center}
\resizebox{0.9\textwidth}{!}{
\includegraphics[width=0.35\textwidth]{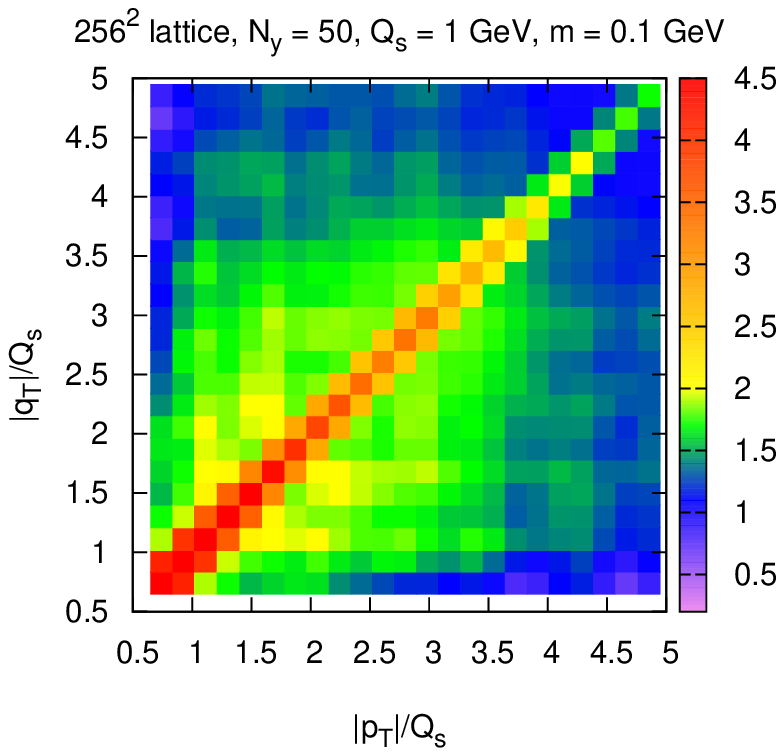}
\rule{1cm}{0pt}
\includegraphics[width=0.35\textwidth]{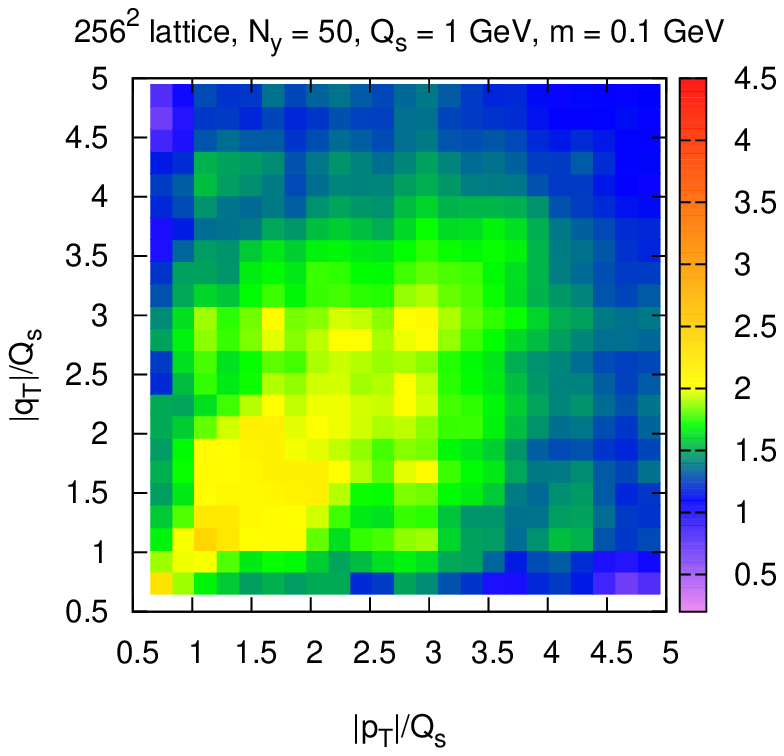}
}
\end{center}
\caption{Numerical evaluation of the two-gluon correlation strength in 
the MV model as a function of the two transverse momenta
$\ptt$ and $\qtt$. Left: the correlation strength
on the ``near side'' $|\varphi_p - \varphi_q| < \pi/2$, right:
the ``away side'' $|\varphi_p - \varphi_q| > \pi/2$.
}\label{fig:kappa2}
\end{figure}

Working with the MV model Gaussian probability distribution
\begin{equation}
W[\rho] = \exp \left[ -\int \ud^2\xt \frac{\rho^a(\xt)\rho^a(\xt)}{g^4\mu^2} \right]
\end{equation}
computing the correlations and the multiplicity distribution
in the linearized approximation
 is a simple combinatorial problem.
Each gluon is produced from two Lipatov vertices 
(see \fig\ref{fig:lipatov} left), 
one in the amplitude and the other in
the complex conjugate. The combinatorial factor is obtained by counting
the different ways of contracting the sources pairwise 
(see \fig\ref{fig:combinat} right).
The dominant contributions are the ones that have, in the dilute limit,
 the strongest infrared divergence 
which is regulated 
by the transverse correlation scale of the problem, $\qs$.
When integrated over the momenta of the produced gluons one 
obtains the factorial moments of the multiplicity, 
which define the whole probability distribution.
 It can be expressed in 
terms of two parameters, the mean multiplicity $\bar{n}$, and a
parameter $k$ describing the width of the distribution.
The result of the combinatorial  exercise is that the 
$q$th factorial moment $m_q$ 
(defined as $\langle N^q \rangle$ minus the corresponding 
disconnected contributions) is
\begin{equation}
 m_q = (q-1)! \,  k \left(\frac{\bar{n}}{k} \right)^q \quad \textrm{with}\quad
k \approx  \frac{  (\nc^2-1)   \qs^2 S_\perp  }{2\pi}
\quad\textrm{and}\quad
\bar{n} =  f_N \frac{1}{\as} \qs^2 S_\perp.
\end{equation}
Here $S_\perp$ is the transverse area of the system
These moments define a \emph{negative binomial} distribution with parameters  
$k$ and $\bar{n}$, known
as a phenomenological observation in high energy hadron and nuclear collisions
already for a long 
time. Also the numerical magnitude of the parameter $k$ obtained from the saturation
scale agrees very well with both pp and AA 
data~\cite{Arnison:1982rm,*Alner:1985zc,*Alner:1985rj,*Ansorge:1988fg,*Adler:2007fj,*Adare:2008ns}.

This calculation predicts that the $k$ parameter should increase with energy, 
unlike what is seen in UA5 data.
Multiplicity fluctuations in proton-proton collisions at lower energies are still 
mostly dominated by the dilute edge of the collision system, causing
a Poissonian nature (i.e. $k\to \infty$) of low energy particle emission. Our
calculation formally assumes $\qs^2 S_\perp \gg 1$, and we expect the 
growing behaior of $k$ with energy to eventually take over at high enough energy.
Some signs of this are
already visible in the LHC $\sqrt{s}=7\textrm{ TeV}$ data\cite{Aamodt:2010pp}.
 
In terms of the glasma flux tube picture this result has a natural  interpretation.
The transverse area of a typical flux tube is $1/\qs^2$, and thus there are 
$\qs^2 S_\perp= N_{\textnormal{FT}}$  independent ones. Each of these radiates
particles independently into $\nc^2-1$ color states in a Bose-Einstein distribution
(see e.g.~\cite{Fukushima:2009er}). A sum of $k \approx N_{\textnormal{FT}} (\nc^2-1)$ 
independent Bose-Einstein-distributions is precisely 
equivalent to  a negative binomial 
distribution with parameter $k$. A numerical evaluation~\cite{Lappi:2009xa} 
of the second moment of the 
distribution, parametrized in terms of
\begin{equation}
\kappa_2(\pt,\qt) = \qs^2 S_\perp \left( 
\frac{\ud^2 N}{\ud^2\pt \ud^2\qt}
-
\frac{\ud N}{\ud^2\pt } \frac{\ud N}{\ud^2\qt }
\right) \bigg/ \frac{\ud N}{\ud^2\pt } \frac{\ud N}{\ud^2\qt }
\end{equation}
is shown in Fig.~\ref{fig:kappa2}. It confirms the expectations 
of~\cite{Gelis:2009wh} that this ratio is of order one and depends only 
weakly on the momenta $\pt,\qt$.

\section{Rapidity dependence}
\label{sec:rapdep}
\begin{figure}
\centerline{
\resizebox{0.9\textwidth}{!}{
\includegraphics[height=5cm]{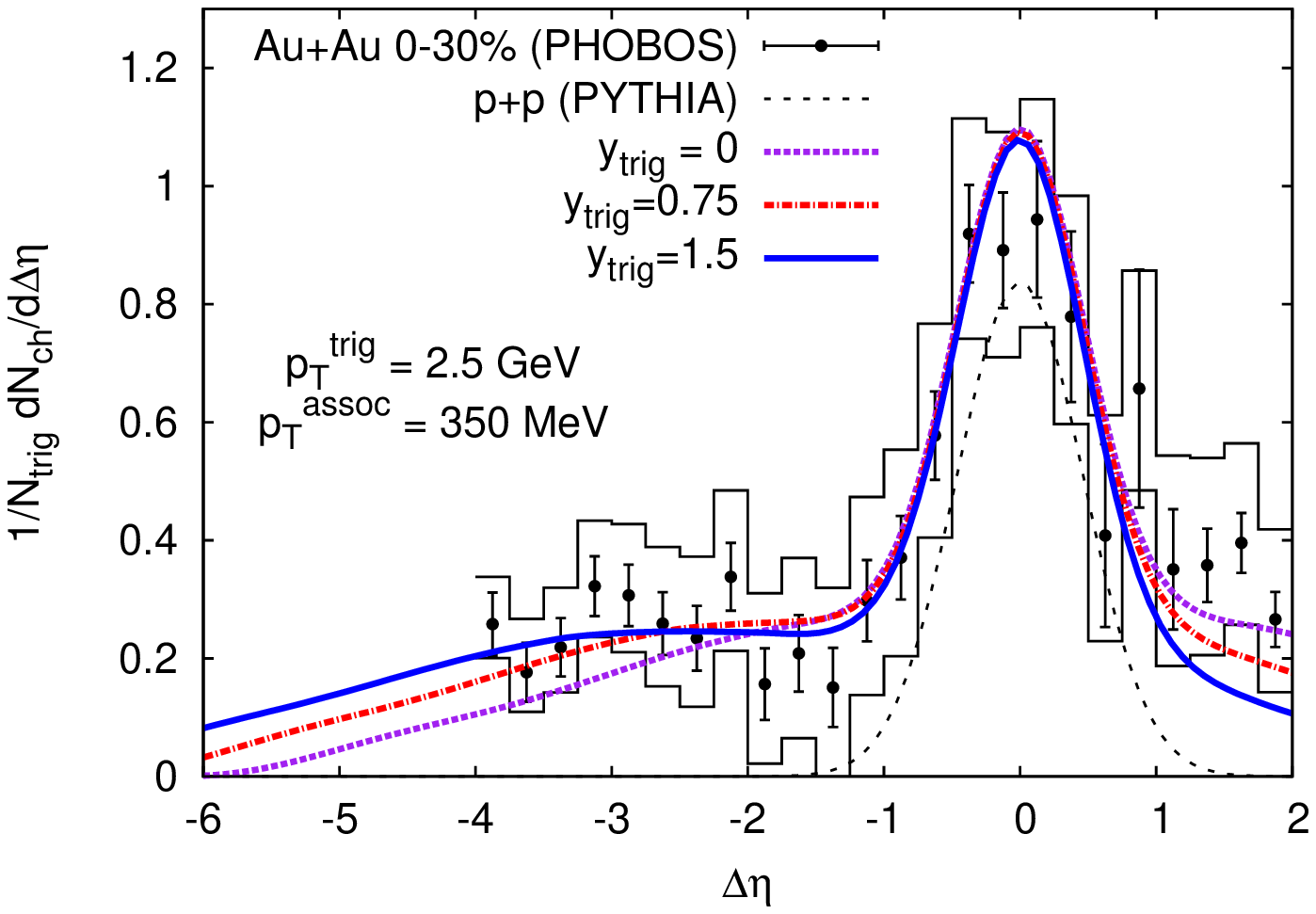}
\includegraphics[height=5.222cm]{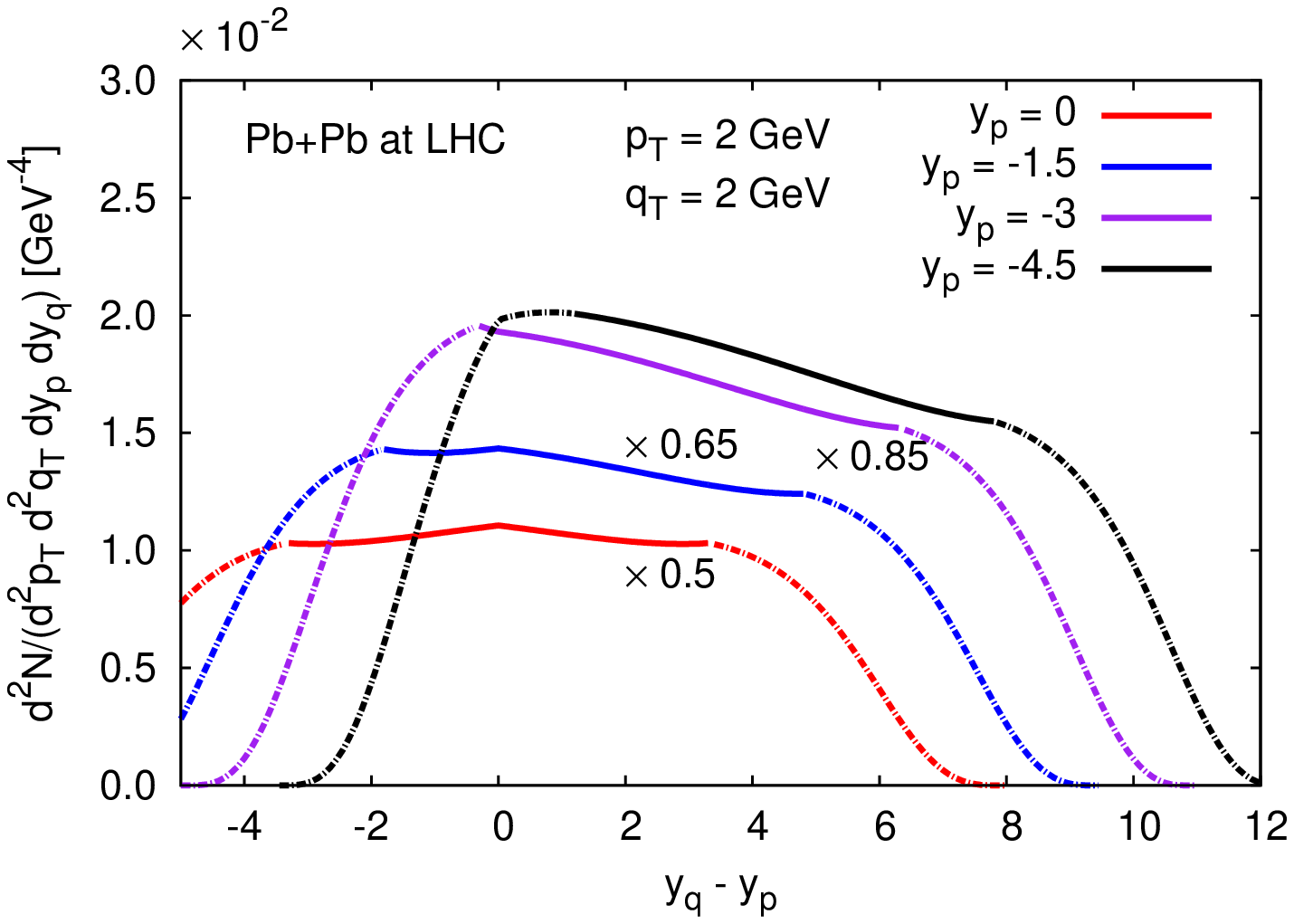}
}
}
\caption{
Left: Comparison of a two-particle correlation computed using 
\eq\nr{eq:double-inclusive-4} supplemented with a short-range correlation 
contribution from PYTHIA with PHOBOS data.
Right: Rapidity correlation at LHC energies $\ktt$-factorization approximation.
Plots from\protect\cite{Dusling:2009ni}.
}\label{fig:rapidity}
\end{figure}

The general discussion of Sec.~\ref{sec:multig} on the different nature of 
multigluon correlations in the ``AA'' case applies also to the rapidity dependence.
Until now we have only been discussing gluon production in a rapidity interval smaller
than $1/\as$. For this we needed only the correlations between the color charges
$\rho(\xt)$ measured at this same rapidity. To understand the rapidity dependence
of the correlations one needs also the correlation between color charges at
different rapidities, $\langle \rho_y(\xt) \rho_{y'}(\yt) \rangle$. 
Also this information is contained in the JIMWLK renormalization group evolution,
at least to  leading $\ln 1/x$ accuracy~\cite{Gelis:2008sz,*Lappi:2009fq}.
An intuitive description of the resulting correlations is provided by the formulation
of JIMWLK as a Langevin equation in the space of Wilson lines formed from the color 
charges. In this picture the evolution proceeds in individual trajectories
along an increasing rapidity.
A first attempt of a realistic estimate of the rapidity dependence 
of two-gluon correlations is performed in Ref.~\cite{Dusling:2009ni}. Evaluating 
the two gluon correlation in a dilute limit in a $\ktt$-factorized 
approximation, but keeping the general  structure resulting from the JIMWLK evolution
leads to the following expression:
\begin{multline}
\label{eq:double-inclusive-4}
C(\p,\q)
=
\frac{\as^{2}}{16 \pi^{10}}
\frac{N_c^2(N_c^2-1) S_\perp}{\da^4\; \pt^2\qt^2}
\\
\times
\bigg\{
\int \ud^2\kt
\Phi_{A_1}^2(y_p,\kt)\Phi_{A_2}(y_p,\pt-\kt)
\Big[
\Phi_{A_2}(y_q,\qt+\kt)
+
\Phi_{A_2}(y_q,\qt-\kt)
\Big]
\\
+
\Phi_{A_2}^2(y_q, \kt)\Phi_{A_1}(y_p,\pt-\kt)
\Big[
\Phi_{A_1}(y_q,\qt+\kt)
+
\Phi_{A_1}(y_q,\qt-\kt)
\Big]
\bigg\}.
\end{multline}
Note the very different structure of this correlation compared to one where
the gluons would be produced from the same diagram for fixed sources. The two 
gluon correlation function is proportional to the product of
\emph{four} unintegrated gluon distributions, with three of them evaluated
at the rapidity of one of the produced gluons and only one at the other. This 
structure is a direct consequence of the nature of JIMWLK evolution. The 
resulting correlation is compared to PHOBOS data in \fig\ref{fig:rapidity}.
The $\ktt$-factorized approximation gives a very inaccurate description of the 
gluon spectrum in the transverse momentum regime $\ptt \sim \qs$ where
the bulk of the particles are produced~\cite{Blaizot:2010kh}. 
Equation~\nr{eq:double-inclusive-4} has also been derived in the 
approximation, true only in the linearized case,
that the unequal rapidity correlation of two color charge densities 
is equal to the unintegrated gluon distribution at the smaller one of 
these rapidities. As of yet there is no calculation of how much this 
approximation is violated in the full JIMWLK evolution. The results presented
in \fig\ref{fig:rapidity} are therefore not the final word on the subject, 
alhough it is reassuring that such a simple approximation seems to agree 
rather well with the experimental result.

\section{Forward-backward multiplicity correlation}

\begin{figure}
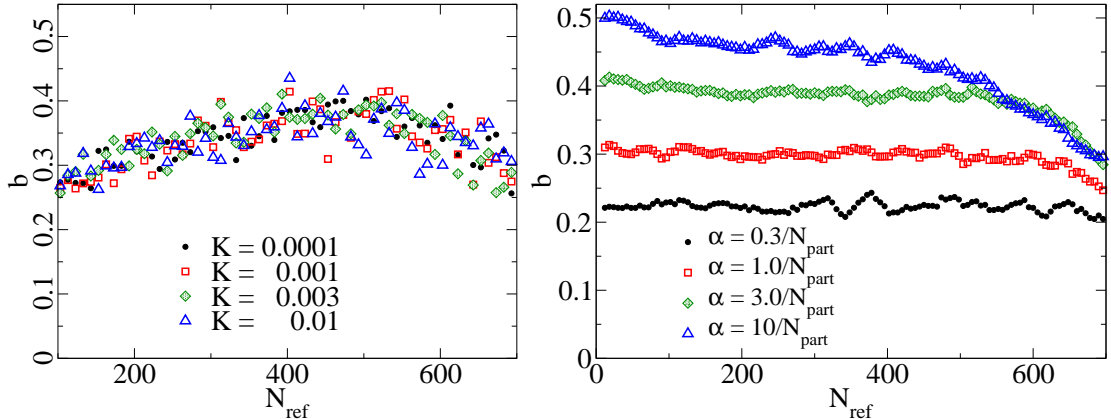
 \label{fig:mcglauber}
\begin{center}
\resizebox{0.9\textwidth}{!}{
\includegraphics[height=5cm]{corrbalfa0.01.eps}
\includegraphics[height=5cm]{varyakcorrbK1.0.eps}
}
\end{center}
\caption{
Simple parametrization of a measured forward-backward correlation 
at fixed reference multiplicity. Left: dependence on 
strength of the long range correlation $K$ is minimal. Right in stead, the
observed correlation strength mostly depends on the short range 
correlation $\alpha$, i.e. the fluctuations of the multiplicities
that are uncorrelated between different rapidities.
}

\end{figure}

Let us conclude by a few remarks on the somewhat puzzling STAR 
data~\cite{Abelev:2009dq} on backward-forward charged hadron multiplicity 
correlations, following the discussion in Ref.~\cite{Lappi:2009vb}.
The reported forward-backward correlation shows a rapid increase as a
 function of centrality, and appears to have a strength which cannot be explained
 by a superposition of pp interactions. 
The measured quantity is usually reported as a correlation coefficient 
between the charged hadron multiplicities measured in a forward and a backward
rapidity bin ($\nf$ and $\nb$)
within the STAR TPC coverage of two units in pseudorapidity:
\begin{equation}
 b = \frac{ \langle \nf \nb \rangle - 
 \langle \nf \rangle \langle \nb \rangle} 
{ \langle \nf^2\rangle
-  \left\langle \nf \right\rangle^2},
\end{equation}
where due to the symmetrical placement of the bins around midrapidity
the variance $ \langle \nf^2\rangle -  \left\langle \nf \right\rangle^2$
is the same for both rapidity bins. 
 If the expectation value here were taken over all
the events in a centrality bin, correlations
would be generated by the  different impact parameters (or numbers of wounded nucleons)
possible within such a  bin\footnote{One expects 
the charged multiplicity to be strongly correlated with impact parameter and if the 
impact parameter itself can have significant variation within a fixed centrality bin,
then spurious correlations whose only origin is the geometry of the 
collision would be generated.}.
This is, however, a too simplified picture of what is actually done in the 
analysis. The nature of the actual quantity reported by STAR is not very explicitly
described in the paper; to understand it one must decipher the inocuous-looking
 statement~\cite{Abelev:2009dq}
``\dots $(\nf,\nb,\nf^2,\nf\nb)$ was obtained on an event-by-event basis as 
a function of the event multiplicity [$\nref$] \dots ''. In other 
words the correlation coefficient is not measured averaging over all the 
events in the bin, but over events with a fixed multiplicity $\nref$ in a third, 
reference, rapidity window, i.e.
\begin{equation}\label{eq:defb}
 b = \frac{ \langle \nf \nb \rangle_{\nref} - 
 \langle \nf \rangle_{\nref} \langle \nb \rangle_{\nref}} 
{ \langle \nf^2\rangle_{\nref} 
-  \left\langle \nf \right\rangle^2_{\nref}} .
\end{equation}
Thus  the measured quantity does not describe the relation between
\emph{two} but \emph{three} correlated multiplicities. For the largest 
separations between the forward and backward rapidity windows the reference
window is between the two, so there is no reason to assume it to be 
less correlated with the $\rmF,\rmB$ windows than these are with each other.
The analysis in Ref.~\cite{Lappi:2009vb} shows that in the limit of a 
\emph{maximal} correlation between multiplicities at different 
rapidities the coefficient ``$b$'' of \eq\nr{eq:defb} reaches a maximum 
value of $1/2$. The results of a more detailed parametrization 
including the fluctuations in centrality and two parameters
describing the  magnitudes of the uncorrelated short range fluctuations
$\sim \alpha$ and the long range correlation $\sim K$ are shown in 
\fig\ref{fig:mcglauber}. The result remains that while the measured 
correlation never exceeds $1/2$, its magnitude depends more on the 
uncorrelated fluctuations than the actual long range correlation. 
The experimental measurements~\cite{Abelev:2009dq} saturate and, 
for the most central bins, even exceed this limit derived on very general
grounds. Although they do seem to point to very strong correlations, 
this inconsistency makes their interpretation in terms of any
microscopic origin, the glasma or something else, of the correlations difficult.

\section{Conclusion}
Most experimental observables do not probe the glasma
initial state of directly, because the system goes through a complicated
time evolution before the hadronization stage. A good candidate for an 
 experimental probe giving direct access to the initial state
is provided by different kinds of
correlation measurements. These have indeed been a focus of 
both experimental and theoretical activity recently. 
We have argued here that the glasma picture of the initial stages
of a heavy ion collision is the natural framework
to understand the origin of these correlations.

\paragraph{Acknowledgements}
The author is supported by the Academy of Finland, contract 126604.

\bibliographystyle{h-physrev4mod2M}
\bibliography{spires}

\end{document}